\documentclass[preprint,showpacs,preprintnumbers,amsmath,amssymb]{revtex4}

\usepackage{amsmath,amssymb}
\usepackage{graphicx}
\usepackage{epsfig}
\usepackage{dcolumn}
\usepackage{bm}

\begin{document}

\title{Weber's Law in Autocatalytic Reaction Networks}

\author{Masayo Inoue$^1$ and Kunihiko Kaneko$^2$}
\affiliation{
$^1$Cybermedia Center, Osaka University, Toyonaka, Osaka 560-0043, Japan \\
$^2$Department of Basic Science, Graduate School of Arts and Sciences, University of Tokyo and ERATO JST,
3-8-1 Komaba, Meguro-ku, Tokyo 153-8902, Japan }
\date{\today}

\begin{abstract}
Biological responses often obey Weber's law, according to which the magnitude of the response depends only on the fold change in the external input. In this study, we demonstrate that a system involving a simple autocatalytic reaction shows such response when a chemical is slowly synthesized by the reaction from a faster influx process. We also show that an autocatalytic reaction process occurring in series or in parallel can obey Weber's law with an oscillatory adaptive response. Considering the simplicity and ubiquity of the autocatalytic process, our proposed mechanism is thought to be commonly observed in biological reactions.
\end{abstract}

\maketitle

Adaptation is ubiquitous in biological systems. One kind of adaptation is perfect adaptation, wherein some state variables of a biological system change in response to a change in the external conditions but slowly come back to their original values \cite{KOSHLAND:1982aa}. Such adaptation is observed in various systems such as signal transduction systems, sensory systems, and neural systems.

Several theoretical studies have investigated such adaptation by using a set of ordinary differential equations, in which certain variables return to their original value independently of the parameter values representing the external condition. Several models for studying perfect adaptation have been proposed and analyzed \cite{Asakura:1984aa, KNOX:1986aa, Barkai:2001aa, Levchenko:2002aa, Erban:2004aa, Inoue:2006aa, Ma:2009aa}.

Responses in a biological system have another ubiquitous characteristic in addition to adaptation: the degree of response is often proportional to the ratio of an external stimulus before and after it is applied, rather than the absolute difference. In other words, the response detects the fold change, which is commonly known as Weber's law. This kind of response was discovered in the field of psychology, wherein it was observed that our sensory response is proportional to the logarithm of the magnitude of an external stimulus.

Recently, Weber's law was also observed to be obeyed by cellular responses in microorganisms \cite{Cohen-Saidon:2009aa, Goentoro:2009aa, Goentoro:2009ab}. Shoval et al. \cite{Shoval:2010aa} further defined fold-change detection (FCD) in a sense stronger than Weber's law: not only the magnitude of the response peak but also the entire relaxation profile over time depends only on the change ratio of the stimuli. In other words, the relaxation profile over time is identical as long as the ratio is constant, irrespective of the absolute magnitude of stimuli. Shoval et al. also developed a theoretical model for such FCD by using a feed-forward gene regulation network. However, it is currently unclear  whether this FCD is ubiquitous, whereas Weber's law itself seems to be quite universal in biological responses.

An adaptive response usually has two components: rapid response to a change and slow relaxation to the original state. Oosawa and Nakaoka confirmed the existence of these distinct timescales in the adaptive response of \textit{Paramecium} to chemotaxis and also the relevance of their existence to chemotaxis \cite{OOSAWA:1977aa} (see also \cite{Inoue:2006aa}). It is therefore important to elucidate the relevance of such a timescale difference to Weber's law for adaptation.

Most biochemical processes involve catalytic reactions, whereas the growth and reproduction of a cell involve an autocatalytic process \cite{Eigen:1978aa, Kauffman:1993aa, Jain:2001aa}. In the present Letter, we show that a simple system with catalytic reactions exhibits an adaptive response according to Weber's law if the autocatalytic process is slow \cite{Inoue:2006aa, KANEKO:1994aa}. Because the autocatalytic process often involves several steps and requires a long time for completion, the ubiquity of adaptive responses obeying Weber's law of a cell can be easily understood from this simple system. We also demonstrate Weber's law in a system of chained or parallel-connected autocatalytic reactions. In addition, we explore oscillatory adaptation with Weber's law and establish a condition for it. Finally, we discuss the relevance of the results to biological responses.

First, we study a simple reaction model of two chemicals $X_0$ and $X_1$ with the following autocatalytic reaction, as introduced in \cite{Inoue:2006aa}: $X_0+X_1 \rightarrow 2 X_1$; the model also includes (a) the synthesis of $X_0$ from an external resource chemical $S$ as $S \rightarrow X_0$ and (b) the degradation of $X_0$ and $X_1$. By representing the concentrations of the two chemicals as $x_0$ and $x_1$ and suitably scaling the time and concentration variables, we get the rate equation as
\begin{equation}
\frac{dx_0}{dt}=S - x_0 x_1 - x_0,  \ \
\frac{dx_1}{dt}=\frac{x_0 x_1 - x_1}{\tau }.
\label{eq_mikk_original}
\end{equation}
The steady state is given by $x_0^* = 1, \ x_1^* = S-1$. According to a linear stability analysis, this state is stable when $S > 1$, i.e., as long as $x_1^*>0$. It should be noted that $x_0^*$ is independent of $S$. The chemical concentration responds to the concentration of the external signal, $S$; when $S$ increases (decreases), $x_0$ increases (decreases) from its steady-state value but later returns to the original value. Thus, $X_0$ always shows an adaptive response, whereas the steady-state concentration of $X_1$ changes with changing $S$.

To examine Weber's law, we study the response of $x_0$ when $S$ changes as $S_0 \rightarrow pS_0 (p>0)$, and calculate the dependence of the peak value of $x_0$ during this change in $S$.  We assume $\tau \gg 1$, which is required to ensure a fast response and slow adaptation. Under the adiabatic limit, $x_1$ changes more slowly than $x_0$ does. Then, during the fast response of $x_0$ to the change in $S$, $x_1$ can be assumed to remain at the steady-state value under the condition of $S=S_0$. Then, the peak value of $x_0$ is obtained from $(dx_0/dt)_{x_0=x_0^{peak}}=0$ by fixing the value of $x_1 $ to $x_1 = S_0 - 1 $. A straightforward calculation gives us $x_0^{peak} = p$. Hence, $x_0$ changes from the original value $x_0^* = 1$ to $x_0^{peak} = p$ and then returns to the original value. This amplitude of the response depends only on $p$, i.e., the ratio of the shift in $S$, and is independent of the $S_0$ value. Thus, our model demonstrates Weber's law under the adiabatic condition.

From a standard linear stability analysis, we get two timescales for this adaptive response: one for response and the other for adaptation. When $S$ is large enough, we can assume that the timescale for adaptation is given just by $\sim \tau$ and is independent of $S$. However, the timescale for response (the peak time) is given by $\sim 1/S$, which is still dependent on $S$. Thus, the temporal profile of the response depends not only on $p$ but also on $S_0$. In this sense, our considered model does not satisfy the condition for FCD as defined by Shoval et al. \cite{Shoval:2010aa}.

\begin{figure}
\begin{center}
\scalebox{0.35}{\includegraphics{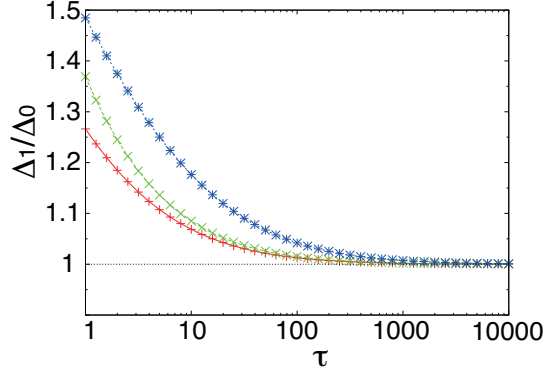}}
\caption{Plot of ratio of maximum response amplitude $\triangle_1 / \triangle_0$(ordinate axis) versus $\tau$(abscissa axis). $\triangle_i = x_0^{peaki} - x_0^*$ with the change $S_i  \rightarrow pS_i$. $\triangle_1 / \triangle_0 =1$ implies the independence of the response amplitude from the $S_i$ value, or Weber's law. Data with three different $(S_0, S_1, p)$ sets are shown: $(S_0, S_1, p)=(5, 50, 2)$ for $+$, $(S_0, S_1, p)=(5, 500, 2)$ for $\times$, and $(S_0, S_1, p)=(5, 50, 10)$ for $\ast$.  }
\label{pic_peak2_NUM2}
\end{center}
\end{figure}

We note that the adiabatic condition need not be rigid in order for Weber's law to be demonstrated reasonably well. Fig.\ref{pic_peak2_NUM2} shows a plot of the peak-value ratio $x_0^{peak1} / x_0^{peak2}$ obtained by multiplying $S$ $p-$fold from $S_0^1$ or $S_0^2$, respectively, as a function of $\tau$. 
The ratio is close to unity; it is independent of the initial $S_0$ as long as $\tau > 10$. In a previous study, we determined that for efficient chemotaxis \cite{Inoue:2006aa}, the ratio of the response time to sense the environmental changes($\tau_s$) to the relaxation time($\tau_a$) is on the order of 100, at several microorganisms such as $\textit{Paramecium}$ \cite{OOSAWA:1977aa} and $\textit{E. coli}$ \cite{BLOCK:1982aa, SEGALL:1986aa}. In the present model, the timescales for response and adaptation are given by $\tau_s \sim 1 / S$ and $\tau_a \sim \tau$, respectively, and the above condition corresponds to $\tau \sim 10$. For this timescale ratio, the peak-value ratio is $ \sim 1.2$ as shown in Fig.\ref{pic_peak2_NUM2}, and thus, Weber's law is approximately obeyed.

Adaptation obeying Weber's law is not restricted to this two-component reaction system; in fact, it is generally observed in a system with autocatalytic reactions with a slower timescale. Here, we study several extensions of autocatalytic reaction networks that show adaptation obeying Weber's law. First, we consider an autocatalytic reaction occurring in series with $N$ chemicals, as $S \rightarrow X_0 \rightarrow X_1\rightarrow ... \rightarrow X_{N-1}$. The rate equations are given by
\begin{align}
dx_0/dt &=(\ \ \ \  S \ \ \ \ \  - \beta_1 x_0 x_1 \ \ \ \  - x_0)/ \tau_0  , \nonumber \\
dx_i/dt &=(\beta_i x_{i-1} x_i - \beta_{i+1} x_i x_{i+1} - x_i)/\tau_i \ \footnotesize{(i \neq 0, N-1)}, \nonumber \\
dx_{N-1}/dt &=(\beta_{N-1} x_{N-2} x_{N-1}   - x_{N-1})/\tau_{N-1} .
\label{eq_mikk_dN}
\end{align}
Then, the steady-state concentrations are obtained as
$x_0^* = S/(\beta_1 x_1^* + 1) , \ x_{i}^* = (\beta_{i+2} x_{i+2}^{*} + 1)/\beta_{i+1} \ (i \neq 0, N-2), and \  x_{N-2}^* = 1/\beta_{N-1} $.
Accordingly, both the steady-state concentrations $x_{N-2}^*$ and $x_{N-2m}^*$ ($m>1$) are independent of the external-signal concentration $S$, and these results show the adaptive response to the change in $S$. Here, we also note that the adaptation alone of the chemical $x_{N-2}$ depends only on the reaction process $X_{N-3}\rightarrow X_{N-2}\rightarrow X_{N-1}$. As long as this chain reaction is autocatalytic and $X_{N-2}$ increases with $S$, $x_{N-2}$ undergoes perfect adaptation. We can modify other reaction processes while retaining the adaptive response of $X_{N-2}$.

First, we study the case of $N=3$ in detail. For simplicity, we use $\beta_i = 1$. Suppose $x_i>0$; the steady state is given by $x_0^* = S/2$, $\ x_1^* = 1$, and $\ x_2^* = (S-2)/2$, which is linearly stable if $S>2$ and $x_1$ shows adaptive response to the change in $S$.

\begin{figure}
\begin{center}
\scalebox{0.5}{\includegraphics{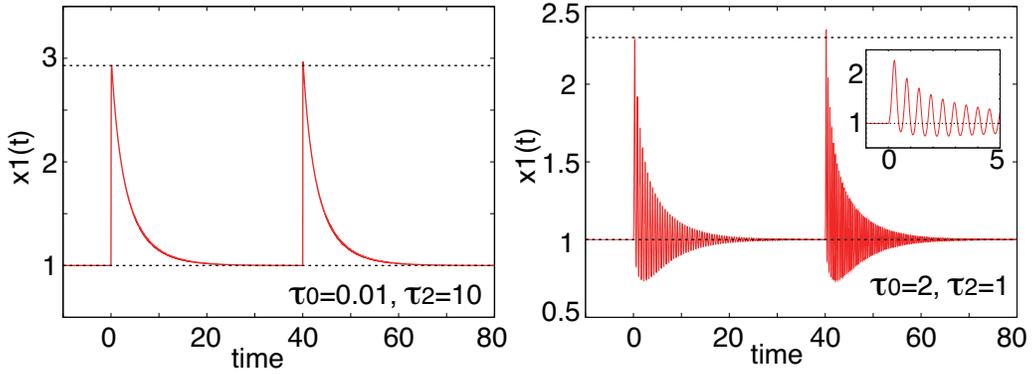}}
\caption{Behaviors of an adaptive variable($x_1$) in the $N=3$ case. Responses corresponding to $S=100 \rightarrow 200$ at $t=0$ and $S=200 \rightarrow 400$ at $t=40$ are shown respectively. (Left) $\tau_0 =0.01, \tau_1 =1, \tau_2 =10$ and (Right) $\tau_0 =2, \tau_1 =1, \tau_2 =1$, zoomed in the inset. }
\label{pic_dynamics3}
\end{center}
\end{figure}

The relaxation process to the state, however, depends on the timescale, as is confirmed from the eigenvalues of the Jacobi matrix (Fig.\ref{pic_dynamics3}). To achieve a normal adaptive response with monotonic relaxation, $\tau_0 \ll \tau_1, \tau_2$ is required (for large $S$, $8\tau_0 <\tau_2$ and $\tau_0/2 < \tau_1/S < \tau_2/32$ are required \cite{cond1}): otherwise, the system would show an adaptive response with damped oscillation.

An adaptive response with damped oscillation has often been observed in experiments, e.g., in \cite{Gregor:2010aa, Ricci:1998aa}. In contrast, theoretical models for such oscillatory adaptation have rarely been explored. Our proposed model provides a simple example of an oscillatory adaptive response.

In an autocatalytic reaction chain as well, Weber's law for responses is obeyed for only a certain range of parameters. Here, we study the conditions for Weber's law in the $N=3$ case as an example. To analyze this case, we again study the system response to the change in $S$ as $S_0 \rightarrow pS_0 (p > 0)$. By setting $\tau_0 \ll \tau_1 \ll \tau_2$, we study the dynamics of $x_1$ and its peak value ($x_1^{peak}$) during an adaptive response. Since the change in $x_0$ is much faster than that in  $x_1$, we can assume that $x_0$ always takes the equilibrium value defined from the present $x_1$ value, and hence, $x_0$ is obtained from $dx_0 / dt = 0$ as $x_0 = p S_0 / (1+x_1) $ for given $x_1$. However, $x_2$ still remains at the equilibrium value under the condition $S=S_0$, i.e., $x_2 = (S_0 - 2)/2$. Using these values, $x_1^{peak}$ that satisfies $(dx_1/ dt)_{x_1 = x_1^{peak} } = 0$ is given by $x_1^{peak} = 2 p -1$. Accordingly, $ x_1^{peak} -x_1^* = 2 (p -1)$ depends only on the ratio of change in $S$ and not on the $S_0$ value itself. Thus, the model given by eq.(\ref{eq_mikk_dN}) with $N=3$ obeys Weber's law under the limits of $\tau_0 / \tau_1 \rightarrow 0$ and $\tau_2 / \tau_1 \rightarrow \infty$.

\begin{figure}
\begin{center}
\scalebox{0.45}{\includegraphics{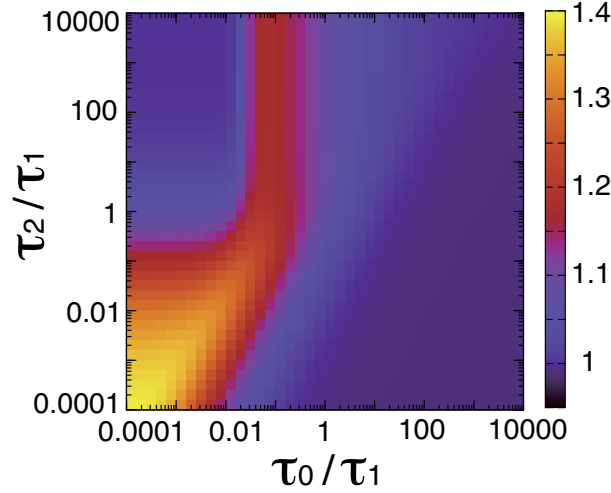}}
\caption{Phase diagram with $\tau_0 / \tau_1$ (abscissa axis) and $\tau_2 / \tau_1$ (ordinate axis). Ratio of the maximum response amplitude, $\triangle_1 / \triangle_0$, is shown. $\triangle_1 = x_1^{peak} - x_1^*$ with $S= 200 \rightarrow 400$ and $\triangle_0$ with  $S= 100 \rightarrow 200$.  }
\label{pic_peak2D_NUM3}
\end{center}
\end{figure}

The above condition for Weber's law is a straightforward extension of the condition derived in the original model with two variables. To examine the validity of the above approximation, we numerically studied the peak-amplitude ratio of $x_1$ under varying $\tau_2$ and $\tau_0$ and constant $\tau_1$. This condition for all $\tau$'s corresponds to the upper-left area in Fig.\ref{pic_peak2D_NUM3}, where Weber's law is confirmed. In this region, $x_1$ shows a monotonous relaxation.

In Fig.\ref{pic_peak2D_NUM3}, however, we find another regime that (approximately) satisfies Weber's law in the lower-right triangle, where $x_1$ shows a damped oscillation and the above adiabatic conditions are not satisfied. Indeed, in this case, Weber's law is well obeyed for large $S$ and $\tau_0$.

\begin{figure}
\begin{center}
\scalebox{0.45}{\includegraphics{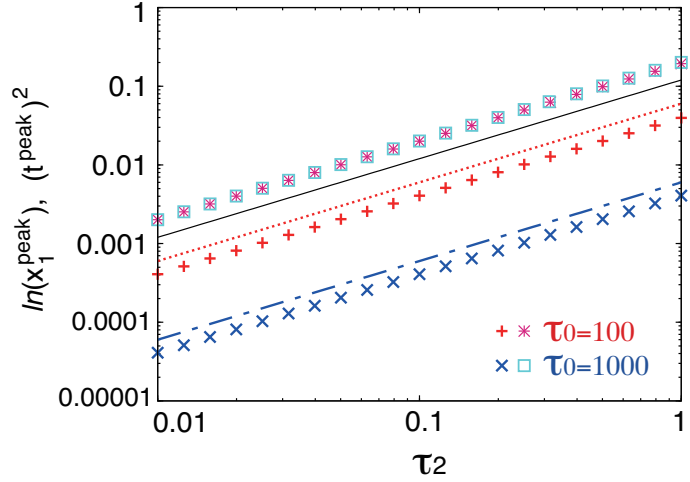}}
\caption{Comparison between results of numerical simulation (points) and approximate calculation (lines). $\ln(x_1^{peak})$ and $(t^{peak} )^2$ are shown as a function of $\tau_2$ with $\tau_1$ fixed to 1. $\ln(x_1^{peak})$ with $\tau_0=100$($+$) and $\tau_0=1000$($\times$) and $(t^{peak} )^2$ with $\tau_0=100$($\ast$) and $\tau_0=1000$($\square$) are taken from the numerical simulation. Eq.(\ref{eq_N3_x1peak}) with $\tau_0=100$ (dotted line) and $\tau_0=1000$ (broken line) and eq.(\ref{eq_N3_tpeak}) (straight line) are also shown. }
\label{pic_peakoscillation_NUM3}
\end{center}
\end{figure}

We now consider the response in the case that $S$ changes as $S_0 \rightarrow pS_0 (p > 0)$ with $\tau_0 \gg \tau_1, \tau_2$ and $S_0 \gg 1$. We solve eq.(\ref{eq_mikk_dN}) with the initial condition $x_0(0)=S_0/2$, $x_1(0)=1$, and $x_2(0)=S_0/2-1$ to obtain the first peak in $x_1$ at $t= t^{peak}$, which is expected to be $t^{peak} \ll 1$ because $S$ is sufficiently large. From the assumption of large $S$ and $\tau_0 \gg \tau_1$, $x_0$ is solved approximately as $x_0(t) \sim \frac{S_0}{2}+ \frac{(p-1)S_0}{\tau_0}t $ for small $t$. Then, we get
\begin{equation}
\frac{d \ln x_1}{dt} \sim \frac{S_0 (p-1)}{\tau_0 \tau_1}t - \frac{S_0}{(S_0 -2) \tau_1}\{ x_2(t) - x_2(0) \}.
\label{eq_d3_lnx1}
\end{equation}
Temporarily neglecting the latter term for small $t$, we get $x_1 \sim \exp \left[ S_0 (p-1) t^2 / (2 \tau_0 \tau_1) \right]$. By substituting this in $d(\ln x_2)/dt =(S_0 -2)(x_1 -1)/(2 \tau_2)$, we get $x_2 = x_2(0) \exp \left[ (p-1) S_0 t^3 / (6 \tau_0 \tau_1 \tau_2) \right]$. Now, with this increase in $x_2$, the second term in eq.(\ref{eq_d3_lnx1}) is no longer negligible, as a result of which the sign of eq.(\ref{eq_d3_lnx1}) changes to negative and $x_1$ stops increasing. Thus, the time of occurrence of the first peak in $x_1$ is estimated from the time when the first term is equal to the second one as 
\begin{equation}
(t^{peak} )^2 \sim 12 \tau_1 \tau_2/S_0 . 
\label{eq_N3_tpeak}
\end{equation} 
By substituting this value in the expression of $x_1$, the peak value of $x_1$ is roughly estimated as
\begin{equation}
x_1^{peak} \sim \exp \left[ \frac{6(p-1) \tau_2}{\tau_0} \right] .
\label{eq_N3_x1peak}
\end{equation}
Hence, the maximum amplitude depends on the change ratio of $S$ and not on $S_0$.
Next, we numerically verified the dependence of the peak amplitude on $\tau$'s from the original equation (eq.(\ref{eq_mikk_dN})). As shown in Fig.\ref{pic_peakoscillation_NUM3}, the above approximations give good estimates for the peak position and explain the validity of Weber's law.

Note that we need to satisfy $t^{peak} \ll \tau_0$ as $\tau_0$ takes the largest value among the three in this area.  Then, we get
\begin{equation}
\frac{\tau_1 \tau_2}{\tau_0^2} \ll \frac{S_0}{12}
\label{eq_N3_boundary},
\end{equation}
which explains the boundary condition in the lower-right area. Note that the peak value itself increases as $\tau_2/\tau_0$ increases. Hence, to achieve an oscillatory response with a large amplitude, the value on the left-hand side of eq.(\ref{eq_N3_boundary}) should preferably be large. Then, an oscillatory response with Weber's law with a relatively large peak will be obtained at the border of the above condition. 

For $N>3$ as well, $x_{N-2}$ shows an adaptive response with both monotonic relaxation and damped oscillation. For the former case, by carrying out similar analysis with $N=3$ case, Weber's law is shown to hold under the condition of $\tau_0, \cdots , \tau_{N-3} \ll \tau_{N-2} \ll \tau_{N-1}$. According to eq.(\ref{eq_mikk_dN}), $x_{N-2m}$ ($m>1$) also shows an adaptive response, and Weber's law for $x_{N-2m}$ is satisfied when $\tau_0, \cdots , \tau_{N-2m-1} \ll \tau_{N-2m} \ll \tau_{N-2m+1}$, i.e., if all processes of influx are fast and only the next runoff process is slow. Weber's law with damped oscillation is also satisfied for large $S$ and $\tau_0$.

We also studied the case of catalytic reaction networks having parallel paths. Two types of extensions are possible. In one extension, many terminal chemicals exist instead of a single $x_1$ in eq.(\ref{eq_mikk_original}), which are reached in parallel from $x_0$ ($S\rightarrow x_0\rightarrow x_1^1, x_1^{2}, x_1^{3}, \cdots x_1^k$). It can be straightforwardly shown that variable $x_0$ shows an adaptive response, since the mean field of all $x_1^j$'s acts as the variable of $x_1$ in the original two-variable case. Moreover, in this type of extension, each catalytic reaction need not be autocatalytic; that is, the reaction from $x_0$ to $x_1^j$ can be catalyzed by $x_1^m$ with $j\neq m$.   Adaptation obeying Weber's law is possible in the adiabatic limit. The other extension includes reaction paths in two or more rows, each of which is a series reaction, as in the case of variable $N$ ($S\rightarrow y_0\rightarrow a_1\rightarrow \cdots a_N \rightarrow y_1$ and $S\rightarrow y_0\rightarrow b_1\rightarrow \cdots b_M \rightarrow y_1$). Here, both an input variable ($y_0$) and an output variable ($y_1$) can show adaptive responses only when there are an odd number of elements in each row ($N$ and $M$). In this case, adaptation obeying Weber's law is again possible, but there usually are more restrictions on the parameter values.

In this Letter, we have demonstrated that a simple autocatalytic reaction process in a system leads to its adaptation that obeys Weber's law. We first confirmed such an adaptation in an autocatalytic reaction of two variables. Coupling of one more variable to the reaction led to oscillatory adaptation obeying Weber's law. It would now be interesting to explore such an adaptive response experimentally. For example, such an oscillatory adaptation was recently observed in cAMP concentration in \textit{Dictyostelium} cells \cite{Gregor:2010aa}. 

Generally, cells undergo autocatalytic reactions to replicate themselves, where complex autocatalytic reactions are often slower than simple catalytic ones. The results of this study suggest that adaptation obeying Weber's law is generally observed in such systems with slow autocatalytic reactions. For example, consider a sequential reaction process for the synthesis of a biopolymer with catalytic activity. With an increase in the sequence length of the polymer, more time would generally be required for its synthesis and degradation; whereas, polymers with a longer sequence have the ability to catalyze these chain reactions. This process is in agreement with the sequential autocatalytic model studied here.

The present model does not show FCD in the strong sense \cite{Shoval:2010aa}. However, the long-term relaxation process depends only on the fold change and not on the absolute value of the external parameter $S$, under the adiabatic condition.

Adaptation or habituation refers to a general property of a biological system to exhibit homeostasis, whereas Weber's law is applicable to a system exhibiting a sensory response over a wide range of environmental conditions. The variation in an external parameter $S$ will often increase with an increase in its value. Then, if the response were just proportional to the difference in the external parameter before and after the input, it would be too sensitive to large $S$ and would not be able to generate a reliable response under external noise. Therefore, a response to fold change in accordance with Weber's law is applicable to sensory systems in general.

In the present study, Weber's law was usually obeyed in an autocatalytic reaction system with timescale differences, without any special design or tuning of the parameters. Because of its simplicity, the present mechanism is expected to have a wide range of applications. It may also offer fresh perspectives on Weber's law in general, including psychological and neural perspectives. Indeed, our proposed model can be regarded simply as a system with self-positive feedback and is not necessarily restricted to chemical reactions.

The authors would like to thank K. Kamino, S. Sawai, and K. Fujimoto for useful discussions. M.I. was partially supported by the JSPS.


\begin{thebibliography}{99}
\bibitem{KOSHLAND:1982aa}
D.E. Koshland, Jr., A. Goldbeter and J.B. Stock, Science \textbf{217}, 220 (1982)
\bibitem{Asakura:1984aa}
S. Asakura and H. Honda, J. Mol. Biol. \textbf{176}, 349 (1984)
\bibitem{KNOX:1986aa}
B.E. Knox, P.N. Devreotes, A. Goldbeter and L.A. Segel, Proc. Natl. Acad. Sci. U.S.A. \textbf{83}, 2345 (1986)
\bibitem{Barkai:2001aa}
N. Barkai, U. Alon and S. Leibler, C. R. Acad. Sci. \textbf{2}, 1 (2001)
\bibitem{Levchenko:2002aa}
A. Levchenko and P.A. Iglesias, Biophysical Journal \textbf{82}, 50 (2002)
\bibitem{Erban:2004aa}
R. Erban and H.G. Othmer, SIAM J. Appl. Math. \textbf{65}, 361 (2004)
\bibitem{Inoue:2006aa}
M. Inoue and K. Kaneko, Phys. Rev. E \textbf{74}, 011903 (2006)
\bibitem{Ma:2009aa}
W. Ma et al., Cell \textbf{138}, 760 (2009)
\bibitem{Cohen-Saidon:2009aa}
Cohen-Saidon et al., Molecular Cell \textbf{36}, 885 (2009)
\bibitem{Goentoro:2009aa}
L. Goentoro and M.W. Kirschner, Molecular Cell \textbf{36}, 872 (2009)
\bibitem{Goentoro:2009ab}
L. Goentoro et al., Molecular Cell \textbf{36}, 894 (2009)
\bibitem{Shoval:2010aa}
O. Shoval et al., Proc. Natl. Acad. Sci. U.S.A. \textbf{36}, 15995 (2010)
\bibitem{OOSAWA:1977aa}
F. Oosawa and Y. Nakaoka, J. Theor. Biol. \textbf{66}, 747 (1977)
\bibitem{Eigen:1978aa}
M. Eigen and P. Schuster, The Hypercycle (Springer, 1979).
\bibitem{Kauffman:1993aa}
S. A. Kauffman, The Origin of Order (Oxford Univ. Press, 1993)
\bibitem{Jain:2001aa}
S. Jain and S. Krishna, Proc. Natl. Acad. Sci. U.S.A. \textbf{98}, 543 (2001)
\bibitem{KANEKO:1994aa}
K. Kaneko and T. Yomo, Physica D \textbf{75}, 89 (1994)
\bibitem{BLOCK:1982aa}
S.M. Block, J.E. Segall and H.C. Berg, Cell \textbf{31}, 215 (1982)
\bibitem{SEGALL:1986aa}
J. Segall, S. Block and H.C. Berg, Proc. Natl. Acad. Sci. U.S.A. \textbf{83}, 8987 (1986)
\bibitem{cond1}
The exact conditions are $x = \frac{\tau_2}{\tau_0} \frac{S}{S-2} >8$ and $\frac{S}{64x} \Big\{ x^2 + 20x - 8 - (x-8) \sqrt{x(x-8)} \} < \frac{\tau_1}{\tau_0} < \frac{S}{64x} \Big\{ x^2 + 20x - 8 + (x-8) \sqrt{x(x-8)} \} $
\bibitem{Gregor:2010aa}
T. Gregor et al., Science \textbf{328}, 1021 (2010)
\bibitem{Ricci:1998aa}
A.J. Ricci, Y.C. Wu and R. Fettiplace, Journal of Neuroscience \textbf{18}, 8261 (1998)
\end{thebibliography}
\end{document}